\documentclass{article}

\usepackage[final]{neurips_2023_ml4ps}

\usepackage{booktabs}
\usepackage{multirow}
\usepackage{adjustbox}

\usepackage[utf8]{inputenc} 
\usepackage[T1]{fontenc}    
\usepackage{hyperref}       
\usepackage{url}            
\usepackage{booktabs}       
\usepackage{amsfonts}       
\usepackage{nicefrac}       
\usepackage{microtype}      
\usepackage{xcolor}         
\usepackage{graphicx}	
\usepackage{amsmath}	
\usepackage{subfig}
\usepackage{booktabs}
\usepackage{multirow}

\title{Gamma Ray AGNs: Estimating Redshifts and Blazar Classification using Neural Networks with smart initialization and self-supervised learning}

\author{%
  Sarvesh Gharat\thanks{https://sarveshvgharat.github.io/} \\
  Centre for Machine Intelligence and Data Science\\
  Indian Institute of Technology Bombay\\
  Mumbai, India \\
  \texttt{sarveshgharat19@gmail.com} 
  \And
  Abhimanyu Borthakur \\
  Department of Electronics and Communication Engineering \\
  Manipal Institute of Technology \\
  Manipal, India 
  \And
  Gopal Bhatta \\
  Janusz Gil Institute of Astronomy\\
  University of Zielona Góra \\
  Zielona Góra, Poland \\
}

\begin{document}

\maketitle

\begin{abstract}
Redshift estimation and the classification of gamma-ray AGNs represent crucial challenges in the field of gamma-ray astronomy. Recent efforts have been made to tackle these problems using traditional machine learning methods. However, the simplicity of existing algorithms, combined with their basic implementations, underscores an opportunity and a need for further advancement in this area. Our approach begins by implementing a Bayesian model for redshift estimation, which can account for uncertainty while providing predictions with the desired confidence level. Subsequently, we address the classification problem by leveraging intelligent initialization techniques and employing soft voting. Additionally, we explore several potential self-supervised algorithms in their conventional form. Lastly, in addition to generating predictions for data with missing outputs, we ensure that the theoretical assertions put forth by both algorithms mutually reinforce each other.
\end{abstract}

\section{Introduction}
Machine learning has emerged as a powerful tool in the field of gamma-ray astrophysics, revolutionizing the way researchers analyze and interpret complex data from gamma-ray detectors and observatories. One of its key applications is in the classification of gamma-ray sources, such as blazars, pulsars, and gamma-ray bursts \cite[see e. g.,][]{agarwal2023classification,narendra2022predicting,finke2021classification}. Particularly, neural networks, can efficiently identify and categorize these sources based on their spectral signatures and temporal behaviors. Moreover, machine learning models aid in the detection of rare and transient gamma-ray events that might otherwise be overlooked using traditional analysis methods \cite{parmiggiani2021deep}. These algorithms excel at pattern recognition and anomaly detection, making them invaluable for identifying subtle variations in gamma-ray emission. Additionally, machine learning techniques contribute to data-driven approaches for source localization and characterization, enhancing our understanding of high-energy astrophysical phenomena \cite[e. g.,][]{taran2023challenging, bhatta2020nature}. Specifically, various machine learning algorithms have been employed in an attempt to solve two of the important problems in gamma-ray astrophysics: estimating the redshift and classifying the undefined class of blazars among observed gamma-ray-emitting AGNs. Addressing these issues using traditional methods is one of the most challenging tasks in astrophysics as they involve acquiring sophisticated optical spectra as well as conducting multi-wavelength observations. In this work, we present our works on these topics utilizing multiple algorithms. The core idea of our method is to employ traditional methods in a smarter way, whether it be for uncertainty quantification or initializing of weights along with employed soft voting. Along with that, for better user experience, we also deploy one of our models on cloud such that someone having no coding experience can use it effectively. Additionally, having trained on same catalog, we also verify whether both these results boil down to same physics.

\section{Methodology}
\subsection{Data Collection and Processing}
Since its launch in 2008, the Fermi Gamma-Ray Space Telescope's onboard instrument called the LAT has been continuously monitoring the high-energy sky \cite{atwood2009large}. In this study, we utilize the Fermi fourth catalog of active galactic nuclei (AGNs) data release 3 (4LAC-DR3; \cite{ajello2022fourth}). The catalog comprises $3407$ individual sources, of which $1806$ sources have known redshifts. On the other hand, the cleaned sample has $3120$ Blazars, out of which $1335$ are BL Lacs and $670$ are identified as FSRQs. Each source is characterized by a set of $41$ different features with randomly missing values reported in this catalog. Following \cite{coronado2023redshift}, we shortlist a set of $24$ features to estimate redshift. Some of the features have a number of missing values which after sufficient exploration are either imputed using most frequent categorical value imputation \cite{lin2020missing} or the targets corresponding to those features are removed leaving us with a total of $1224$ data points out of which $90\%$ are used for training. On the other hand, for the classification purpose we proceed with $7$ features such that no source has to be removed as doing so may result in a more imbalanced dataset, affecting the results and hence, limiting the scope of the study. A few of these features further undergo a log transformation so that the large values of parameters don't explicitly distort the learning of the model even after suitable normalization.

\subsection{Architecture and Training}
In the first part of our work, we propose a multi-layer perceptron (MLP) with a single hidden layer containing $64$ neurons for redshift estimation. To prevent overfitting, we introduce a dropout rate of $0.25$ in the hidden layer and employ the Rectified Linear Unit (ReLU) activation function for non-linearity. In the output layer, we utilize the softplus activation function, and the chosen loss function is Mean Absolute Error (MAE). This initial model, referred to as the "frequentist" model, treats parameters as point estimates.

To address uncertainty in redshift estimation, we apply variational inference, utilizing two different estimators. Variational inference approximates the posterior distribution of model parameters given observed data which is often intractable for complex models like neural networks \cite{shridhar2019comprehensive, jospin2022hands}. VI introduces an approximating distribution, typically a tractable one, and assigns a prior distribution to model parameters which in our case is a Gaussian distribution. The prior is further updated using Bayes' rule as data is observed, resulting in the posterior distribution. Unlike traditional neural networks, variational inference offers a more meaningful measure of uncertainty. We implement these models using TensorFlow Probability \cite{tensorflow2015-whitepaper} and Keras \cite{chollet2015keras}, utilizing DenseFlipout and DenseReparameterization layers \cite{wen2018flipout, kingma2013auto}. Further the uncertainty is quantified by evaluating each sample $1000$ times, and capturing the variance in these predictions. The Bayesian model's prediction is obtained as the mean from these iterations and finally the standard 3-sigma rule is applied to establish confidence intervals for redshift values, with flexibility for different confidence levels depending on the allowed tolerance.

The second part of this study explores $5$ different classification models including Artificial Neural Networks with $3$ different initialisation techniques and $2$ different self-supervised learning techniques, specifically autoencoders \cite{chen2023context} and contrastive classification \cite{henaff2020data, oord2018representation, liu2023devil}. Autoencoders are adapted for self-supervised learning, training the encoder to extract meaningful data representations, which are then used as inputs for classification. The contrastive classification model focuses on capturing similarities and dissimilarities between data samples, specifically between positive (FSRQ) and negative (BL Lac) samples. The extracted features from this model are incorporated into a standard classifier.

These models provide a comprehensive approach to both AGN redshift estimation and Blazar classification, addressing challenges in weight initialization through supervised and unsupervised pretraining, and leveraging self-supervised learning techniques to enhance classification performance. 

\section{Results and Discussion}




\begin{table*}
\caption{Performance Analysis of Regressor: RMSE and Correlation Coefficient Comparison between Previous Approaches and Our Proposed Models.}    
\centering
\begin{adjustbox}{max width=\textwidth, center, scale=1}
\label{tab:performanceanalysis} 
\begin{tabular}{lllllllllllll}
\toprule
& \cite{dainotti2021predicting} & \cite{narendra2022predicting} & \cite{coronado2023redshift} &  Frequentist & Flipout & Reparameterization \\ \midrule
RMSE & 0.432-0.438 & 0.458 & 0.46 & 0.415 & 0.406 & 0.438 \\ \midrule
CC & 0.704-0.718 & 0.74 & 0.71 & 0.784 & 0.777 & 0.778 \\ \bottomrule
\end{tabular}
\end{adjustbox}
\end{table*}

\begin{table*}
\caption{Performance Analysis of Classifier: A detailed comparison of all our models along with some of the state of the art literature.}    
\centering
\begin{adjustbox}{max width=\textwidth, center, scale=1}
\label{tab:perf_summary} 
\begin{tabular}{llllllll}
\toprule
Model & Accuracy & \multicolumn{3}{c}{BLLac} & \multicolumn{3}{c}{FSRQ} \\ 
\cmidrule(lr){3-5} \cmidrule(lr){6-8}
 & & Precision & Recall & F1-score & Precision & Recall & F1-score \\
\midrule
Bias Initialization with Soft Voting & 0.930 & 0.938 & 0.965 & 0.951 & 0.909 & 0.847 & 0.877 \\
Supervised Greedy Pretraining & 0.915 & 0.963 & 0.915 & 0.939 & 0.818 & 0.915 & 0.864 \\
Unsupervised Greedy Pretraining & 0.886 & 0.941 & 0.894  & 0.917 & 0.773 & 0.864 & 0.816 \\
SSL Autoencoder Pretext & 0.891 & 0.941 & 0.901 & 0.921 & 0.785 & 0.864 & 0.823\\
SSL Contrastive Classification  & 0.886 & 0.941 & 0.894 & 0.917 & 0.773 & 0.864 & 0.816\\
\cite{agarwal2023classification} & 0.938 & 0.956  & 0.952 & 0.954 & 0.902 & 0.910 & 0.906\\
\cite{tolamatti2023classification} & 0.96 & 0.96 & 0.96 & 0.96 & 0.96 & 0.96 & 0.96 \\
\bottomrule
\end{tabular}
\end{adjustbox}
\end{table*}

\begin{table*}
  \caption{Redshift Prediction Summary Statistics: Here the mean prediction refers to the mean of all samples of known and unknown redshift respectively}
  \label{tab:summary_statistics}
  \centering
  \begin{adjustbox}{max width=\textwidth, center, scale=1}
  \begin{tabular}{lccccccc}
    \toprule
    \multirow{2}{*}{Method} & \multicolumn{3}{c}{Known Redshift samples} & \multicolumn{3}{c}{Unknown Redshift samples} \\
    \cmidrule(lr){2-4} \cmidrule(lr){5-7}
    & Mean Prediction & Range & $\sigma$ & Mean Prediction & Range & $\sigma$ \\
    \midrule
    Frequentist Model & 0.559 & 0.04 - 1.99 & 0.372 & 0.455 & 0.07 - 1.77 & 0.258 \\
    Flipout Estimator & 0.581 & 0.027 - 2.11 & 0.382 & 0.415 & 0.0251 - 1.71 & 0.246 \\
    Reparameterization Estimator & 0.526 & 0.0004 - 1.82 & 0.332 & 0.393 & 0.0089 - 1.47 & 0.207 \\
    \bottomrule
  \end{tabular}
  \end{adjustbox}
\end{table*}

\begin{figure*}%
 \centering
 \subfloat[Variational Inference (Flipout Estimator)]{\includegraphics[scale = 0.185]{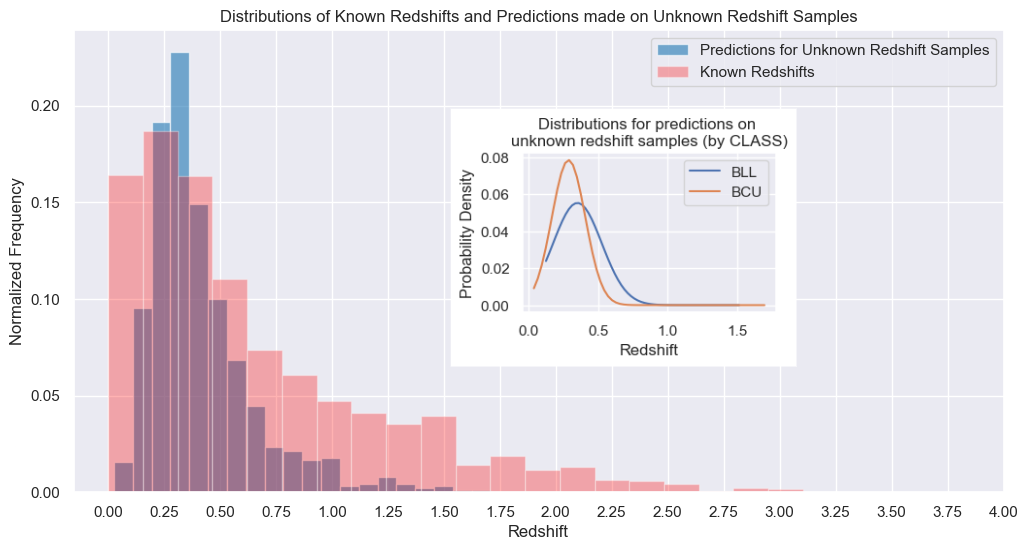}\label{fig:BNN1_redshift}} \quad
 \subfloat[Variational Inference (Reparameterized Estimator)]{\includegraphics[scale = 0.185]{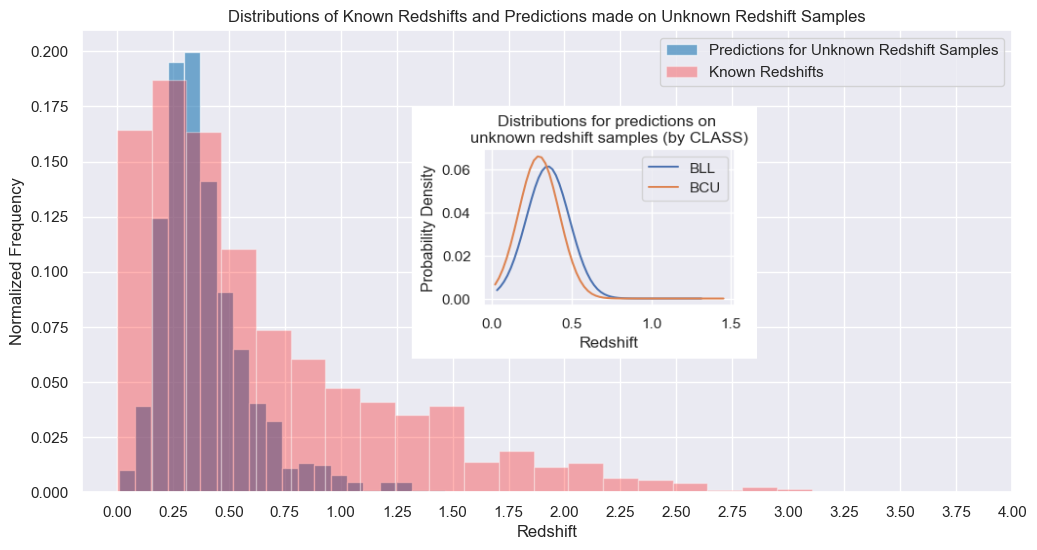}\label{fig:MC1_redshift}}\quad
 \subfloat[Frequentist model]{\includegraphics[scale = 0.185]{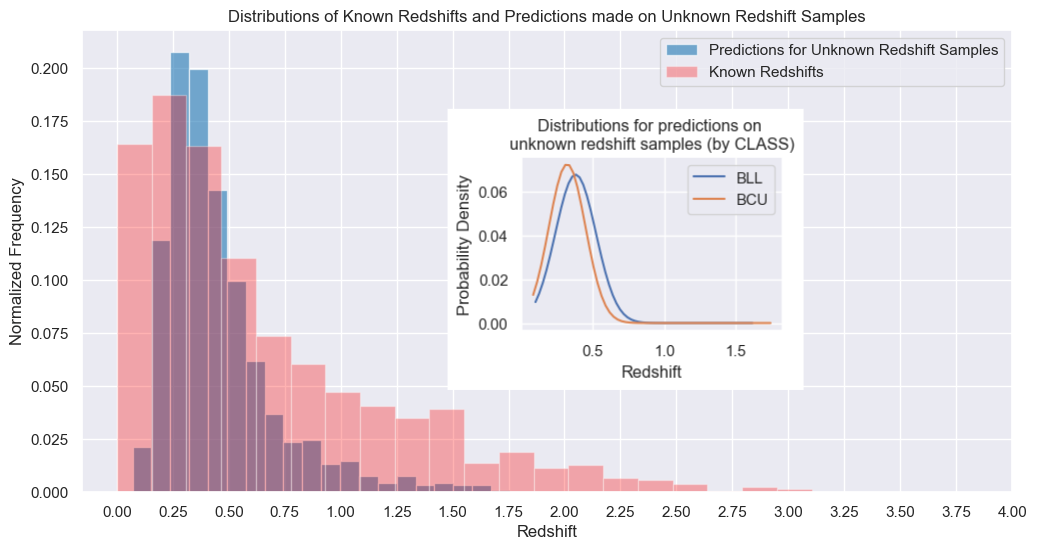}\label{fig:frequentist_redshift1}}%
 \caption{Comparison between Known Redshift Samples and Predictions on Unknown Redshift samples using Histograms. Subplots show the distribution of the redshift values for the predictions made on the unknown redshift samples, disaggregated by the "CLASS" feature.Here, we represent only those classes with more than 50 samples}%
 \label{fig:redshift_histogram1}
\end{figure*}

For a couple of years, there have been a few papers addressing the problem of redshift estimation, however none of them account for uncertainty. Considering the techniques and the kind of data available for classification, there seems be a larger scope for exploration in this domain. Besides that, the proposed work in this study provides some optimal results along with a lower number of parameter making the model simpler but smarter and easy to deploy. In Table \ref{tab:performanceanalysis}, we compare the results from our study for redshift estimation along with the existing literature corroborating the optimal nature of our results along with uncertainty quantification. In a similar fashion, we achieve a result for Blazar classification that is comparable to it's counterparts as shown in Table \ref{tab:perf_summary}. As seen from Table \ref{tab:performanceanalysis}, both RMSE and the correlation coefficient are significantly better than the existing algorithms. In addition to optimal results, the proposed algorithm also employs uncertainty helping the user to know with the confidence of the predictions with possible range. Talking about classification, although from Table \ref{tab:perf_summary}, it may seem that \cite{tolamatti2023classification} performs the best, they have included a large number of features which limits their study to only $112$ BCUs, owing to a high percentage of missing values. On the other hand \cite{agarwal2023classification} makes just use of $7$ features similar to our study, however their results are almost similar to ours with almost $7$ times more parameters making the number of floating point operations (FLOPs) to $5056$ as compared to $672$ in ours. Considering that similar results are achieved with lesser parameters, allowing us to speed up the prediction by $7$ times just through smarter initialization, we deploy the algorithm on Streamlit (\href{https://bcu-classification-ml.streamlit.app/}{https://bcu-classification-ml.streamlit.app/}) and AWS (\href{http://13.239.10.157:8501/}{http://13.239.10.157:8501/}).

Next, we make use of our algorithm to give an estimate on redshifts along with associated uncertainty with appropriate confidence. The summary of the same can be seen in Table \ref{tab:summary_statistics}. From Figure \ref{fig:redshift_histogram1}, we see that for a majority of targets the redshifts have a significantly lower value resulting in a majority of targets being BL Lac objects. The same can be confirmed from the classification results achieved. We see that out of a total of $1115$ BCU samples, $820$ are BL Lacs and $295$ are FSRQs indicating that both these methods employed for different purposes boil down to stating a single claim - that there are more BLLs than FSRQs in the catalog.

\section{Conclusion}
In this study, we encompass multiple algorithms to identify the redshifts along with the associated uncertainty and classifying the Blazars into BL Lac and FSRQs. The employed algorithms, though simple, incorporate Bayesian principles and intelligent weight initialization making them a potent alternative to the existing algorithms while providing improved performance. Additionally, due to the effectiveness of the classifier, we deploy the corresponding algorithm on Streamlit, simplifying user access and interaction. Next, along with the predicted values for unknown redshifts and classes of BCUs, we also verify if both the algorithms support each other in theoretical claims. We see that the regressor claims to have a huge number of lower redshifts in the catalog indicating that there are more BLLs compared to FSRQs. The same claim has been verified via the classifier wherein it confirms the presence of more BLLs in the catalog.

\section{Data Availability}
The contained study is distributed over two manuscripts. For the complete versions and their GitHub repositories, please refer to \cite{gharat2024estimation} and \cite{bhatta2024gamma}.

\bibliographystyle{mnras}
\bibliography{neurips_2023}

\end{document}